# Enhancing the Stretchability of Two-Dimensional Materials through Kirigami: A Molecular Dynamics Study on Tungsten Disulfide

K. Dey[1,†], S. Shahriar[1], M. A. R. Anan[2], P. Malakar[3],
M. M. Rahman[4] and M. M. Chowdhury[5,6,‡]

[1]Department of Mechanical Engineering, Bangladesh University of Engineering and Technology, Dhaka-1000, Bangladesh
[2]Department of Mechanical Engineering, The University of New Mexico, Albuquerque, NM 87131, United States
[3]Department of Materials Science and Engineering, University of Michigan, Ann Arbor, Michigan, 48109, United States
[4]School of Electrical and Computer Engineering, Purdue University, West Lafayette, Indiana 47907, United States
[5]Department of Electrical and Computer Engineering, University of British Columbia, Vancouver, British Columbia V6T 1Z4, Canada
[6]Quantum Matter Institute, University of British Columbia, Vancouver, British Columbia V6T 1Z4, Canada

**Abstract**

In recent years, the 'kirigami' technique has gained significant attention for creating meta-structures and meta-materials with exceptional characteristics, such as unprecedented stretchability. These properties, not typically inherent in the original materials or structures, present new opportunities for applications in stretchable electronics and photovoltaics. However, despite its scientific and practical significance, the application of kirigami patterning on a monolayer of tungsten disulfide ($WS_2$), a van der Waals material with exceptional mechanical, electronic, and optical properties, has remained unexplored. This study utilizes molecular dynamics (MD) simulations to investigate the mechanical properties of monolayer $WS_2$ with rectangular kirigami cuts. We find that, under tensile loading, the $WS_2$ based kirigami structure exhibits a notable increase in tensile strain and a decrease in strength, thus demonstrating the effectiveness of the kirigami cutting technique in enhancing the stretchability of monolayer $WS_2$. Additionally, increasing the overlap ratio enhances the stretchability of the structure, allowing for tailored high strength or high strain requirements. Furthermore, our observations reveal that increasing the density of cuts and reducing the length-to-width ratio of the kirigami nanosheet further improve the fracture strain, thereby enhancing the overall stretchability of the proposed kirigami patterned structure of $WS_2$.

---

[†] koushik.dey.buet@gmail.com
[‡] mchowdhury@ece.ubc.ca

**1 Introduction**

Two-dimensional (2D) materials, including graphene, hexagonal boron nitride (h-BN), phosphorene, and transition metal dichalcogenides (TMDs), have garnered significant attention due to their favorable physical, chemical, optical, and electrical properties and diverse applications [1-4]. Beyond their exceptional properties, the capacity to stack these materials to construct desired heterostructures without introducing interface-induced nonidealities arising from lattice mismatches not only renders 2D materials an ideal platform for exploring novel physical phenomena but also for fabricating advanced heterostructured devices..However, despite such a promise, a lot of these 2D materials are limited by their low bending stiffness [5, 6], making them more susceptible to bending compared to bulk materials. To address this limitation, kirigami, an innovative technique for advanced 3D nanofabrication through the cutting of thin sheets, offers a promising solution [7]. Kirigami enables the creation of meta-structures and meta-materials with unique properties, such as a negative Poisson's ratio, accurate shape morphing, tunable auxetics, super-stretchability, buckling, and multistability through the modification of certain materials or structures [8-11]. The exceptional stretchability and flexibility of kirigami metastructures have been harnessed to develop systems, equipment, macromaterials, and microstructures [12], revolutionizing the research and applications of 2D materials in various fields [13].

Recent research endeavors have focused on leveraging kirigami to enhance the pliability of 2D materials. Molecular dynamics (MD) simulations have been employed to investigate the mechanical properties of kirigami-modified 2D materials, including $MoS_2$ [14, 15], graphene [16, 17], and h-BN [18], phosphorene [19] under various conditions. In particular, the application of kirigami engineering was first explored with pristine graphene which has led to a significant improvement in fracture strain properties, with fracture strains of graphene monolayer kirigami [16] almost three times larger than those of pristine monolayer graphene [20]. Similarly, the kirigami techniques have led to substantial enhancements in the mechanical properties of $MoS_2$ (fourfold increase in yield strain and sixfold increase in fracture strain), h-BN (three to five times increase in fracture strain), and phosphorene (almost twofold increase in fracture strain) compared to their pristine counterparts [21] [22] [23].

Tungsten disulfide ($WS_2$), a two-dimensional material, has gained prominence in various applications such as gas sensing applications, optical modulators, solid and dry film lubricants and as self-lubricating composite materials [24]. It exhibits promising potential in the fields of nanoelectronics, spintronics, and optoelectronics, owing to its desirable properties such as a 2-eV energy gap between bound and free states, low heat conduction, and strong spin-orbit interaction [25]. Notably, $WS_2$ demonstrates better tribological properties and heat resistance than $MoS_2$, making it a viable alternative for similar applications [24]. Literature suggests that kirigami techniques greatly enhance the strechabillity of monolayer $MoS_2$, surpassing the improvements reported for kirigami graphene by Qi et al. [16]. By extending the kirigami approach to enhance the stretchability of monolayer pristine $WS_2$, exciting possibilities emerge for its utilization in stretchable electronics and photovoltaics.

The aim of this study is to extend investigations upon previous successful applications of the concept of kirigami [11] to enhance the stretchability of 2D materials, e.g., graphene, $MoS_2$, phosphorene and hBN. Our specific focus is to investigate the efficacy of the kirigami concept in improving the stretchability of $WS_2$, a distinct 2D material. In this study, rectangular kirigami cuts are introduced to pristine monolayer $WS_2$ nanosheets. Then, molecular dynamics (MD) simulations will be employed to analyze the impacts of kirigami cuts on the flexibility of monolayer pristine $WS_2$ under uniaxial tensile loading. A comprehensive comparison of stress-strain responses between kirigami $WS_2$ and its monolayer pristine counterpart will be presented. Furthermore, the influence of chiral orientations (armchair (AC) and zigzag (ZZ)), overlap ratio and nanosheet length-to-width ratio on the mechanical parameters of the kirigami structure, including Young's modulus (YM), ultimate tensile stress (UTS), and fracture strain, will be investigated.

## 2 Models and simulation methods

The $WS_2$ kirigami sheets were created by applying a constant number of rectangular kirigami cuts (four inner cuts and five pairs of edge cuts) to a monolayer pristine $WS_2$ nanosheet. A schematic top view of the kirigami $WS_2$ structure and the related geometric parameters is shown in Figure 1 along with the AC and ZZ edges (side views) in the model. The main geometric parameters are the nanosheet length $L_0$, width b, height of each inner cut w, height of edge cuts 0.5w, the width of each inner cut c, and the distance between successive cuts d. The overlap ratio, $\alpha$, is defined as the overlapping cut length to the nanosheet length, given by $\alpha = (w - 0.5b)/L_0$. Similarly, the ratio of the overlapping width to the nanosheet length is defined as $\beta = (0.5d - c)/L_0$. Additionally, the length-to-width ratio of the nanosheet is denoted as $\gamma = L_0/b$. To manipulate the overlap ratio, $\alpha$, we vary the value of w while keeping b and $L_0$ constant. Similarly, $\beta$ is adjusted by changing the value of d while maintaining c and $L_0$ constant. As reported in literature [14, 16], while $\alpha$ describes the geometry perpendicular to the tensile loading direction, $\beta$ describes the geometry parallel to the tensile loading direction. In our investigation, we vary the value of $\gamma$ by changing $L_0$ while keeping the number of kirigami cuts constant, which leads to an altered value of d. In this study, we considered kirigami for both AC and ZZ edges. A representative AC $WS_2$ kirigami consisting of a number of, N ~18984 atoms with a nanosheet length $L_0$ ~300 Å, width b ~300 Å, height of each inner cut w ~ 265 Å, width of each inner cut c ~ 19 Å, and distance between successive cuts d ~65 Å is shown in Figure 1.

The MD simulations on the kirigami were conducted employing the widely used open-source code LAMMPS [28], developed by Sandia, while incorporating the Stillinger-Weber (SW) potential specifically tailored for $WS_2$ as proposed by Jiang et al. [26]. The calibration of this potential was based on the phonon dispersion phenomena of single-layer $1H-WS_2$, which encompasses both out-of-plane and in-plane vibrational motions[29]. So, the SW potential is expected to effectively capture out-of-plane deflections involving changes in angles and rotations.

In the simulation procedure, after energy minimization, NVE, NPT and NVT ensembles were applied to the system for 50 ps, 100 ps and 20 ps respectively for equilibration at a timestep of 1 fs. Then, uniaxial tensile strain was applied to the two opposing edges of the nanosheet with a $10^9$ $s^{-1}$ strain rate maintaining periodic boundary conditions in every direction. The stresses were estimated using virial stress theorem [30]. The structures were created using atomsk [31] and atomic visualizations were done with the OVITO package [32].

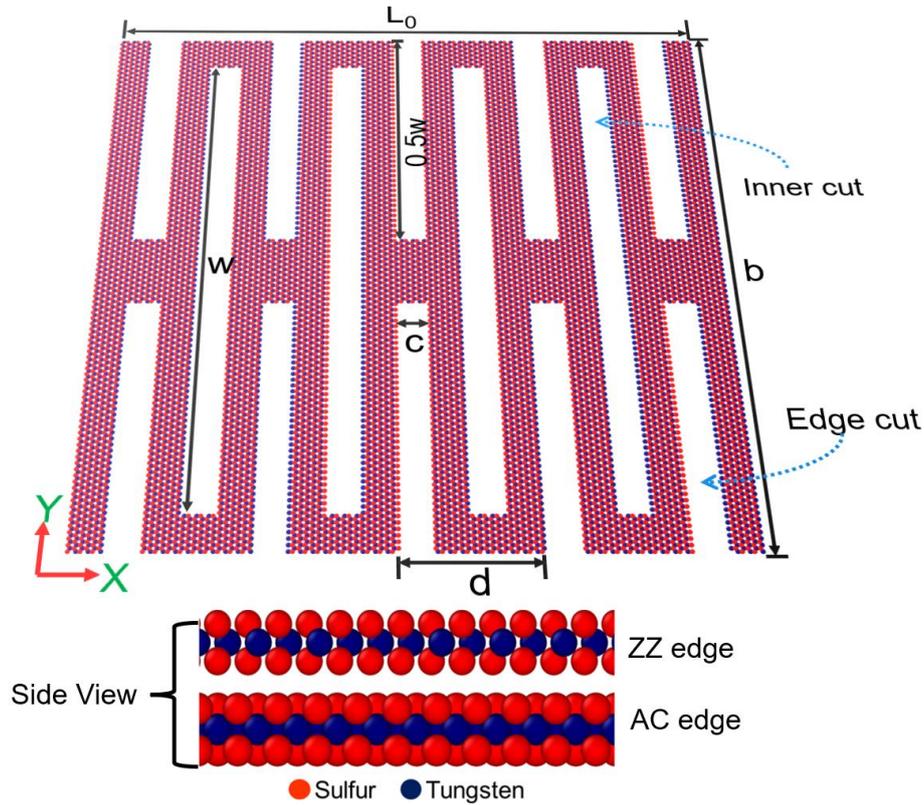

Figure 1: Snapshot of the top view of monolayer kirigami WS$_2$ nanosheet labeled with essential geometric parameters. The Armchair (AC) and Zigzag (ZZ) edges are shown from the side views of the nanosheet.

## 3 Results and discussion

We initially conducted MD simulations on a monolayer pristine WS$_2$ nanosheet (300 Å × 300 Å) to obtain the stress-strain response. Figure 2(a) shows how the stress in pristine WS$_2$ varies with applied uniaxial tensile strain along the AC and ZZ edges. Similar to the previous studies [27, 33], we have observed anisotropic mechanical characteristics when applying uniaxial tensile strain along the AC and ZZ edges. Our study shows that pristine WS$_2$ can undergo tensile strain of up to approximately 17%, with slightly higher stretching observed along the ZZ edge compared to the AC edge.

At the point of peak stress, we have found the fracture strength to be 12.61 N/m along the AC edge, slightly higher than that observed along the ZZ edge. To determine Young's modulus, we employed Hooke's law and linearly fitted the stress-strain curve at low strain levels (<0.2%). The calculated Young's modulus values are 90.16 N/m for the AC edge and 88 N/m for the ZZ edge. For the specific nanosheet size of 54.2 Å × 62.6 Å, which has also been used in previous study [34] our findings are consistent with the literature, as summarized in Table 1. The agreement between our findings and the existing literature underscores the accuracy of our simulator.

**Table 1: Mechanical characteristics of monolayer pristine $WS_2$ under uniaxial tensile loading**

| Material Direction | Fracture Strain | Fracture Strength (N/m) | Young's Modulus (N/m) |
|---|---|---|---|
| AC (Current Study) | 18.1% | 13.77 | 114 |
| AC [34] | 18.63% | 14 | 115 |
| ZZ (Current Study) | 22% | 13.73 | 111.8 |
| ZZ [34] | 23% | 13.9 | 113.2 |

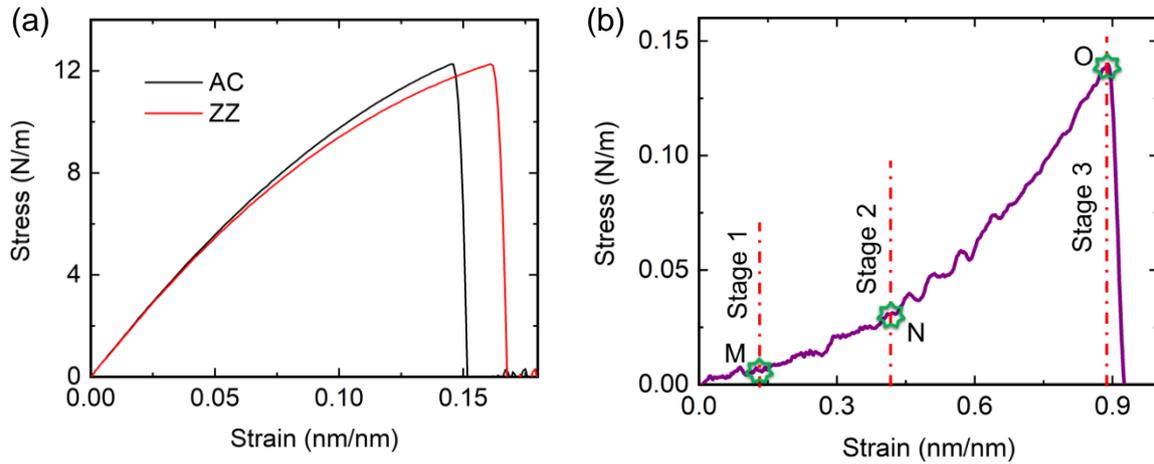

Figure 2: (a) Stress- strain responses of monolayer pristine $WS_2$ nanosheet along AC and ZZ edges. Fracture strain obtained along ZZ edge is higher than that of obtained along AC edge. (b) Stress-strain curve of kirigami engineered monolayer $WS_2$ nanosheet ($\alpha = 0.38$) with 3 deformation stages when applied uniaxial tensile strain along AC edge.

To closely examine the effect of the kirigami cutting method on the mechanical properties of monolayer $WS_2$, we analyze the deformation stages of a kirigami $WS_2$ nanosheet. Figure 2(b) illustrates the three deformation stages with applied tension at the AC edge of the model, where the kirigami structure had an overlap ratio of $\alpha = 0.38$, and the simulation was conducted at a temperature of 300 K. The ending of deformation stages 1, 2 and 3 are marked with points M, N and O, respectively.

In stage 1, a slight increment in stress (~0.005 N/m) is observed at the onset of deformation as the strain reaches up to 15%. Progressing into stage 2, the stress rises to ~0.029 N/m with further strain increase from 15% to 43%. Transitioning to stage 3, a substantial surge in stress (~0.14 N/m) occurs as the strain continues to rise from 43% to 90%. The deformation stages observed in our study are consistent with those reported in a previous study of hBN kirigami by Gamil et al. [19].

Next, we delve into a more detailed examination of the three deformation stages observed here. Figure 3 illustrates the atomic visualization of the deformation process of a kirigami $WS_2$ nanosheet ($\alpha = 0.38$) at various timesteps of 0.15 ns, 0.55 ns, 0.85 ns and 0.90 ns. As mentioned earlier, the kirigami nanosheet is subjected to uniaxial tensile strain along the AC edge (x-direction). The color coding in the figure represents per atom stress, ranging from 0 to $1.5 \times 10^6$ GPa·nm².

In deformation stage 1, the kirigami structure expands by approximately 7%, with the inner cuts (denoted by w in Figure 1) exhibiting slight elongation along the loading direction (Figure 3(a)). The stress increment in stage 1 is negligible (~0.005 N/m). In this stage, the deformation mechanism primarily involves elastic bond stretching, while no flipping or rotation of the $WS_2$ kirigami sheet is observed. This is in contrast to previous studies on graphene kirigami [16], where rotation and flipping were observed in the initial stage. Hanakata et al. [14] reported that the absence of such behavior could be attributed to the higher elastic bending modulus of the structure under investigation compared to monolayer graphene [35].

Transitioning to stage 2, we have observed substantial out-of-plane deformation of the rectangular cuts under the applied tension. Simultaneously, the cuts undergo a transformation, gradually deforming into convex like patterns (i.e., middle section is thicker than edges) [11, 36]. Within this stage, the stress exhibits a notable increase from 0.005 N/m to 0.029 N/m as the strain progressively rises from 7% to 43%. The emergence of the convex pattern becomes clearly visible after 0.55 ns, as shown in Figure 3(b).

In stage 3, as the structure undergoes further stretching along the loading direction, the out-of-plane deformation of the inner cuts gradually diminishes. As the applied load causes the kirigami sheet to stretch further, it becomes nearly flat. Notably, the convex pattern persists even at 0.85 ns of deformation (Figure 3(c)). The stress within the structure reaches a significant level (~0.14 N/m), resulting in stress concentration at the mechanical joints of the cuts, as indicated by the color code in Figure 3. Eventually, fracture occurs at a strain of 90% as shown in Figure 3(d). The crack point, denoted by O in Figure 2(b), corresponds to the location of maximum stress (~0.14 N/m) which is also indicated by the dashed red circles in Figure 3. Within a narrow strain range of approximately 2%, the kirigami structure experiences complete rupture following the initiation of the crack.

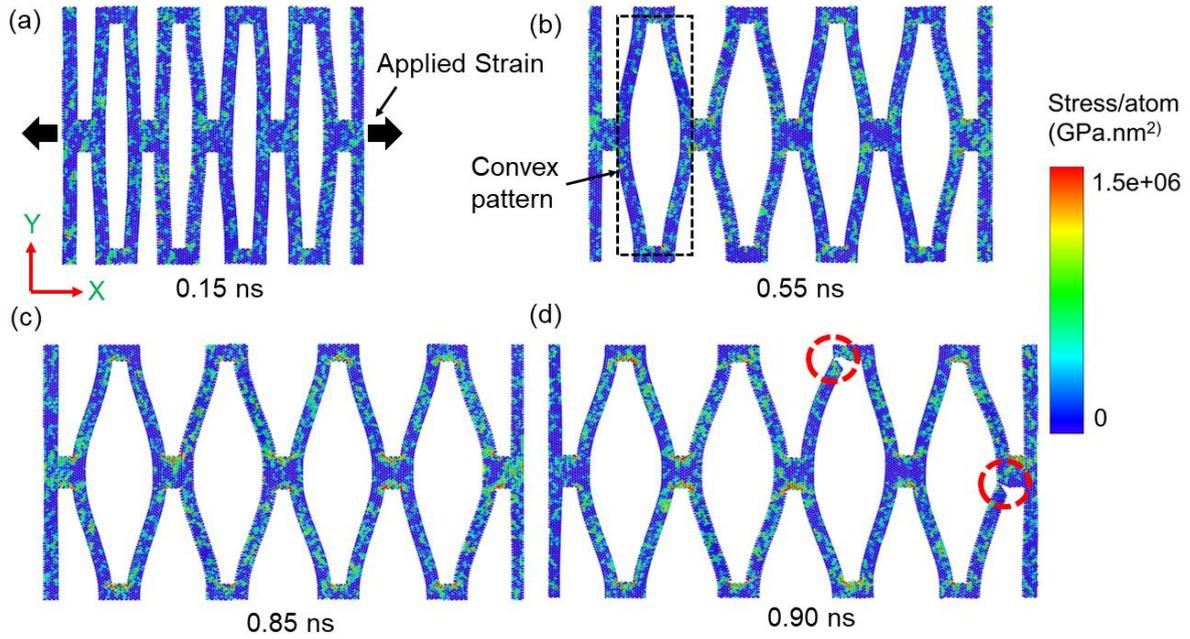

Figure 3: Atomic visualization of the deformation process of a kirigami WS$_2$ nanosheet ($\alpha$ = 0.38) at various timestamps (0.15 ns, 0.55 ns, 0.85 ns, 0.90 ns) at a temperature of 300 K. The kirigami nanosheet is subjected to uniaxial tensile strain along the AC edge (*x*-direction). At 0.90 ns, cracks (indicated in red circles) are observed both at internal and external portion of rhomboid structure. The color coding in the figure represents per atom stress, ranging from 0 to 1.5×10$^6$ GPa·nm².

Figure 4 shows the influence of the overlap ratio ($\alpha$) on the stress-strain behavior of kirigami-WS$_2$ models along the AC and ZZ edges. Increasing the cut length, denoted as w in Figure 1, leads to higher values of $\alpha$, and vice versa. Figure 4 indicates that by controlling the overlap ratio, we can effectively manipulate the flexibility and strength of the kirigami nanosheet. Under the tensile loading along the AC edge, as $\alpha$ increases from 0 to 0.38, the fracture strain increases from ~10.1% to ~88.3%, while the UTS decreases from ~0.545 N/m to ~0.152 N/m. Similarly, along the ZZ edge, increasing $\alpha$ from 0 to 0.38 leads to an increase in fracture strain from ~8.6% to ~88.8%, accompanied by a decrease in UTS from approximately ~0.484 N/m to ~0.147 N/m. The ZZ edge exhibits a relatively lower UTS, but a slightly higher fracture strain compared to that of along the AC edge. The variation in stress response can be attributed to the bond structures [37] of $WS_2$.

$WS_2$ features a trigonal-prismatic crystal structure known as h-$WS_2$, where a layer of tungsten atoms is sandwiched between two layers of sulfur atoms. In the ZZ edge, the S-W-S and S-W-S' bonds are densely packed compared to the AC edge, where S and S' belong to different groups on the top or bottom side [38]. The bond angles along the ZZ edge can stretch further before fracture under tensile loading [34]. This distinction may provide a possible explanation for the slightly higher fracture strain observed when loaded along the ZZ edge as compared to the AC edge.

The kirigami structure we have studied demonstrates a significant trade-off between stress and strain. Higher flexibility is accompanied by a lower fracture stress point, and vice versa. This trade-off is evident in the trends observed in Figure 4(c), illustrating how an increased overlap ratio results in a decrease in UTS of the kirigami nanosheet. Conversely, higher overlap ratios significantly enhance the fracture strain.

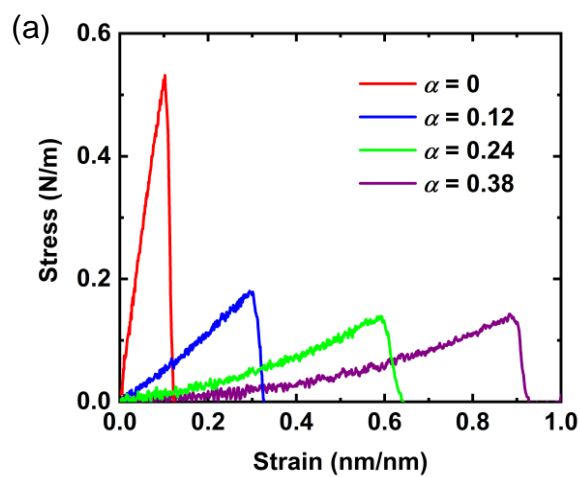

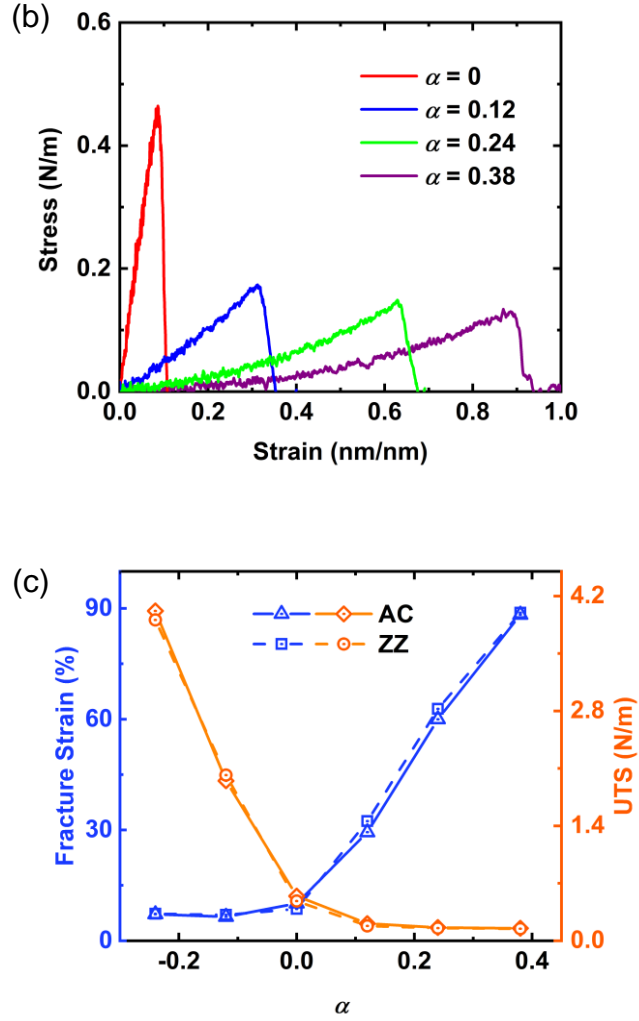

Figure 4. Stress-strain curves of kirigami $WS_2$ with different overlap ratios ($α$) and constant $β = 0.045$, $γ = 1$ under uniaxial tensile strain at 300 K temperature; (a) along the armchair (AC) edge; (b) along the zigzag (ZZ) edge. (c) Variation of Fracture Strain and UTS with overlap ratio, $α$.

The deformation mechanism observed at different $α$ values can be explained as follows. $α = 0$ represents the configuration where the edge and inner cuts begin to overlap (Figure 1). Han et al. [18] reported that when $α$ is less than 0, the edge and inner cuts do not overlap, severely constraining or even inhibiting out-of-plane deformation at lower $α$ values. In this scenario, the cuts can be regarded as atomic line defects, which can significantly reduce the strength and strain of the structure. Smaller values of $\alpha$ correspond to higher strength, approaching that of the cut-free nanosheet. Similar kind of observations were made for $\alpha$ values of -0.24, -0.12 and 0. Hence, our analysis indicates that at lower overlap ratios ($\alpha \leq 0$), the desired advantages of the kirigami patterned cut, such as high stretchability, cannot be achieved. This phenomenon has been observed in previous studies on various materials, such as kirigami graphene [16], $MoS_2$ [14], hBN [18] and phosphorene [19].

On the other hand, our results demonstrate that higher overlap ratios enable a sufficient trade-off between strength and flexibility along both the AC and ZZ edges. Increasing the $\alpha$ value from 0.12 to 0.38 resulted in a decrease in stress from 0.21 N/m to 0.152 N/m, while the fracture strain increased from 29.4% to 88.3%. We have observed slightly higher fracture strain (32.4% to 88.8%) along the ZZ edge, while the fracture stress decreased from 0.182 N/m to 0.147 N/m, indicating a similar pattern as along the AC edge. Han et al. [18] reported that when $\alpha$ is greater than 0, the flipping and rotation mechanism at stage 2 and 3 becomes prominent. This mechanism significantly contributes to the extensive stretching and flexibility of the kirigami configuration. With an increasing overlap between the edge and interior cuts, the stretchability of the kirigami $WS_2$ structure progressively increases.

The observed effect of the overlap ratio in this study aligns with previous investigations on kirigami graphene [16], $MoS_2$ [14], hBN [18] and phosphorene [19]. In a study exploring the effects of various parameters on the mechanical properties of kirigami $MoS_2$ [14], the authors reported an increase in fracture strain by a factor of 5 compared to its pristine counterpart. In our study, we have observed an increase in fracture strain by a factor of approximately 7, providing evidence that kirigami significantly enhances flexibility.

Next, we investigated the influence of $\beta$ on the flexibility of the kirigami nanosheet. The value of $\beta$ is changed by altering $d$ while all other geometric factors remain unchanged.

The results presented in Figure 5(a) demonstrate that, under the tensile loading along the AC edge, an increase in $\beta$ from 0.009 to 0.045 corresponds to a decrease in fracture strain from approximately 128.7% to 88.3%, while the UTS shows an increase from about 0.079 N/m to 0.152 N/m. Likewise, along the ZZ edge, an increase in $\beta$ from 0.009 to 0.045 is accompanied by a decrease in fracture strain from approximately 133.8% to 88.8%, along with an increase in UTS from around 0.091 N/m to 0.147 N/m. This behavior can be attributed to the fact that, when $\beta$ values are higher, the density of cuts decreases, resulting in kirigami structures that exhibit characteristics almost similar to pristine structures. Consequently, this leads to lower fracture strain and higher UTS. Because of the same reason, we observe almost similar scenario along ZZ directions when $\beta$ is higher (Figure 5(a)). These observations are consistent with findings reported in prior studies for other kirigami-cut materials [14, 16, 18].

To examine the impact of the length-to-width ratio ($\gamma$), we varied $\gamma$ while keeping $\alpha$ constant. Our findings reveal that, under the tensile loading along the AC edge, as $\gamma$ decreases from 2.33 to 0.5, the fracture strain (Figure 5(b)) increases from ~27% to ~164%, while the UTS decreases from ~0.25 N/m to ~0.105 N/m. Similarly, along the ZZ edge, decreasing $\gamma$ from 2.33 to 0.5 leads to an increase in fracture strain from ~30% to ~195%, accompanied by a decrease in UTS from ~0.245 N/m to ~0.077 N/m. In our investigation, we increased the value of $\gamma$ while keeping the number of kirigami cuts constant, resulting in a higher value of $d$ and a lower density of cuts. This reduction in cut density caused the kirigami structure to display characteristics more akin to pristine structures, consistent with previous study shown in Figure 5(a). As a result, the fracture strain decreased and the UTS increased with respect to $\gamma$ (Figure 5(b)). Additionally, the differences in behavior between the AC and ZZ edges became less pronounced as $\gamma$ increased (Figure 5(b)), owing to the same underlying mechanism. Our findings indicate that, in addition to $\alpha$, we can also maximize the failure strain by increasing the density of the cuts (decreasing $\beta$) and reducing the length-to-width ratio (decreasing $\gamma$).

We simulated pristine monolayer WS$_2$ nanosheets with varying length and width to investigate the impact of $\gamma$ on fracture strain and UTS (Figure 5c). In contrast to the kirigami structure, we did not observe significant changes in fracture strain or UTS in the case of monolayer pristine WS$_2$ when varying $\gamma$ values. Decreasing $\gamma$ from 2.33 to 0.5 led to an 11.6% increase in fracture strain and a 5.5% increase in UTS along the AC edge. Similarly, along the ZZ edge, we observed a 9.7% increase in fracture strain and a 3.56% increase in UTS. On the other hand, for the kirigami structure (Figure 5(b)), decreasing $\gamma$ from 2.33 to 0.5 resulted in a more than 500% increase in fracture strain and a 69% decrease in UTS. These tradeoffs and the broad range of values provide suitable engineering opportunities (Figure 5(a) and 5(b)), allowing for the selection of an appropriate $\beta$ and $\gamma$ values based on the desired flexibility and strength requirements.

Furthermore, Figure 5(b) illustrates that the fracture strain differences along the AC and ZZ edges are approximately 30% when $\gamma$ is around 0.3, which is approximately 15 times higher than that of the pristine structure shown in Figure 5(c). Figures 5(a) and 5(b) demonstrate that, unlike the pristine structure, by selecting appropriate values of $\beta$ and $\gamma$, one can significantly modify the mechanical anisotropy of the kirigami structure under investigation.

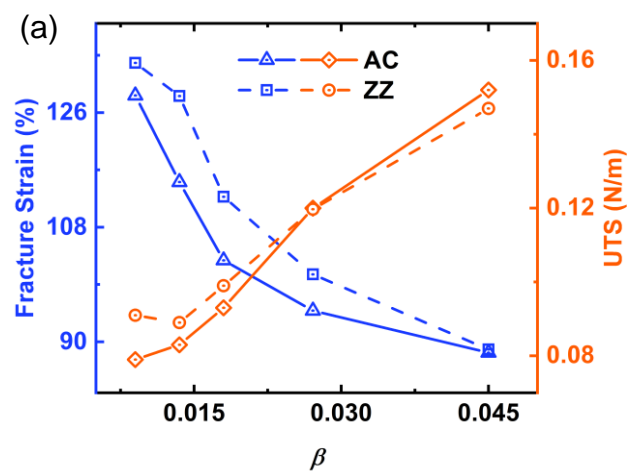

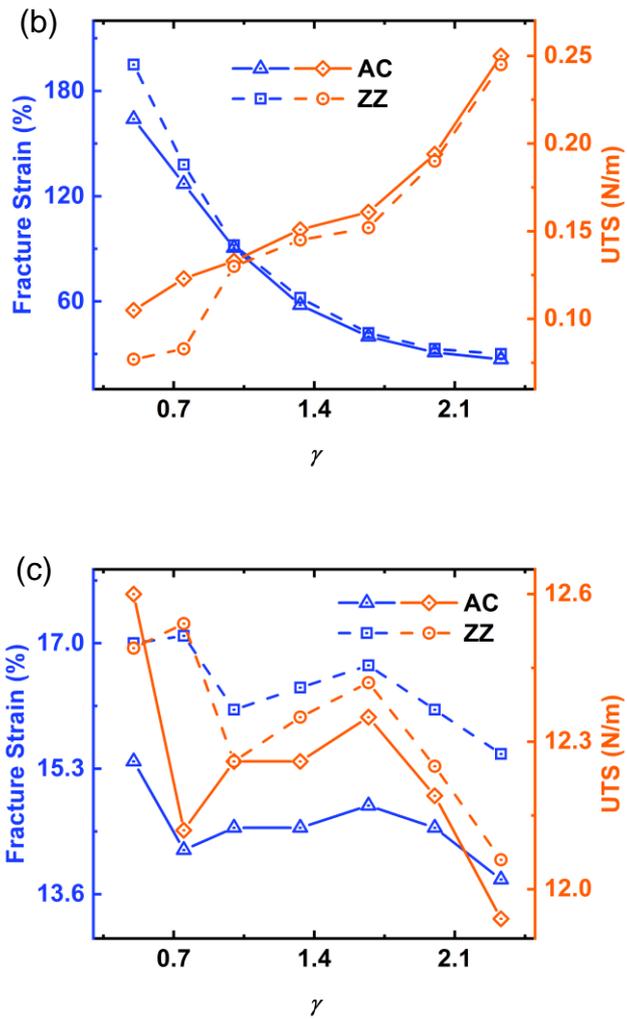

Figure 5: Influence of density of the cuts (*β*) and length-to-width ratios (*γ*) on fracture strain and UTS of kirigami WS$_2$ nanosheet under uniaxial tensile strain along AC and ZZ directions. (a) Variation of Fracture Strain and UTS with respect to density of cuts, *β*; Variation of Fracture Strain and UTS with respect to length to width ratio, *γ* in (b) kirigami WS$_2$ (c) pristine WS$_2$ nanosheets.

**4 Conclusions**

This study employs molecular dynamics simulations to investigate the mechanical properties of monolayer $WS_2$ with rectangular kirigami cuts. Under tensile loading, our findings have revealed a notable increase in tensile strain and a decrease in strength along both the AC and ZZ edges in the kirigami structure, thus demonstrating the effectiveness of the kirigami cutting technique in enhancing the strechabillity of monolayer $WS_2$. Similar to other kirigami-cut 2D materials, we have observed that kirigami patterning enables significant out-of-plane deformations in monolayer $WS_2$, thereby facilitating tensile stretching of the monolayer. The ultimate fracture strain can undergo a nearly six-fold increase, while the corresponding fracture stress exhibits a decrease by a factor of 90 when compared to the pristine monolayer $WS_2$ sheet. Furthermore, we have observed that, by increasing overlap ratio, the flexibility or stretchability of the $WS_2$ based kirigami structure could be increased, allowing for the tuning of high strength or high ductility requirements. Our findings indicate that we can also maximize the failure strain by increasing the density of the cuts (decreasing $β$). Furthermore, for the kirigami structure, decreasing the length-to-width ratio ($γ$) 4.6 times, while keeping the number of kirigami cuts constant, resulted in a more than 500% increase in fracture strain and a 69% decrease in UTS. These tradeoffs and the broad range of values provide suitable engineering opportunities, allowing for the selection of appropriate $α$, $β$ and $γ$ values based on the desired mechanical anisotropy, flexibility and strength requirements. Overall, the results from our study demonstrate the effectiveness of the kirigami cutting technique in enhancing the stretchability of monolayer $WS_2$, thus highlighting the potential of this technique in the fields of stretchable electronics and photovoltaics.

**Acknowledgements**

The authors of this paper would like to acknowledge Multiscale Mechanical Modeling and Research Network (MMMRN) for their technical assistance in carrying out the research.